\documentclass[aps,prb,twocolumn,superscriptaddress,showpacs]{revtex4}

\usepackage{graphicx}% Include figure files
\usepackage{dcolumn}% Align table columns on decimal point
\usepackage{bm}% bold math

\begin{document}

\title{Magnetic phase transitions and magnetoelectric coupling of
GdFeO$_3$ single crystals probed by low-temperature heat
transport}

\author{Z. Y. Zhao}
\affiliation{Hefei National Laboratory for Physical Sciences at
Microscale, University of Science and Technology of China, Hefei,
Anhui 230026, People's Republic of China}

\author{X. M. Wang}
\affiliation{Hefei National Laboratory for Physical Sciences at
Microscale, University of Science and Technology of China, Hefei,
Anhui 230026, People's Republic of China}

\author{C. Fan}
\affiliation{Hefei National Laboratory for Physical Sciences at
Microscale, University of Science and Technology of China, Hefei,
Anhui 230026, People's Republic of China}

\author{W. Tao}
\affiliation{Hefei National Laboratory for Physical Sciences at
Microscale, University of Science and Technology of China, Hefei,
Anhui 230026, People's Republic of China}

\author{X. G. Liu}
\affiliation{Hefei National Laboratory for Physical Sciences at
Microscale, University of Science and Technology of China, Hefei,
Anhui 230026, People's Republic of China}

\author{W. P. Ke}
\affiliation{Hefei National Laboratory for Physical Sciences at
Microscale, University of Science and Technology of China, Hefei,
Anhui 230026, People's Republic of China}

\author{F. B. Zhang}
\affiliation{Department of Materials Science and Engineering,
University of Science and Technology of China, Hefei, Anhui
230026, People's Republic of China}

\author{X. Zhao}
\affiliation{School of Physical Sciences, University of Science
and Technology of China, Hefei, Anhui 230026, People's Republic of
China}

\author{X. F. Sun}
\email{xfsun@ustc.edu.cn}

\affiliation{Hefei National Laboratory for Physical Sciences at
Microscale, University of Science and Technology of China, Hefei,
Anhui 230026, People's Republic of China}

\date{\today}

\begin{abstract}

The low-temperature thermal conductivity ($\kappa$) of GdFeO$_3$
single crystals is found to be strongly dependent on magnetic
field. The low-field $\kappa (H)$ curves show two ``dips" for $H
\parallel a$ and only one ``dip" for $H \parallel c$, with
the characteristic fields having good correspondence with the
spin-flop and the spin-polarization transitions. A remarkable
phenomenon is that the subKelvin thermal conductivity shows
hysteretic behaviors on the history of applying magnetic field,
that is, the $\kappa(H)$ isotherms measured with field increasing
are larger than those with field decreasing. Intriguingly, the
broad region of magnetic field ($\sim$ 0--3 T) showing the
irreversibility of heat transport coincides with that presenting
the ferroelectricity. It is discussed that the irreversible
$\kappa(H)$ behaviors are due to the phonon scattering by
ferroelectric domain walls. This result shows an experimental
feature that points to the capability of controlling the
ferroelectric domain structures by magnetic field in multiferroic
materials.

\end{abstract}

\pacs{66.70.-f, 75.47.-m, 75.50.-y}
%66.70.-f Nonelectronic thermal conduction and heat-pulse propagation in solids
%75.47.-m Magnetotransport phenomena; materials for magnetotransport
%75.50.-y Studies of specific magnetic materials

\maketitle

\section{Introduction}

Low-temperature heat transport is an important physical property
of solids and is useful for probing many kinds of elementary
excitations, such as phonons, electrons, magnons and spinons, etc.
Thermal conductivity ($\kappa$) is strongly dependent on the
statistical laws of these excitations and their transport
properties, which are directly related to the nature of the ground
states of materials. For example, the temperature dependence of
very-low-$T$ thermal conductivity can directly show the purely
phononic transport, the pairing symmetries of superconductors, the
Fermi-liquid state of metals and the nature of spin liquid in
quantum magnets.\cite{Berman, Hussey, Taillefer, Proust,
Yamashita1, Yamashita2} When the scattering between different
types of quasiparticles is significant, the low-$T$ heat transport
can also be an effective way to detect such couplings; in
particular, the spin-phonon or magnon-phonon couplings in
insulating magnetic materials are sometimes easily revealed by the
measurement of the magnetic-field dependence of thermal
conductivity.\cite{Sun_PLCO, Sun_GBCO, Sharma, Wang} In cases of
magnetic excitations either transporting heat or strongly
scattering phonons, the magnetic phase transitions, including the
changes of either the ground state or the spin structure, can be
sensitively probed by the thermal conductivity
measurements.\cite{Sun_LSCO, Sun_YBCO, Paglione, Sologubenko,
Sun_DTN, Takeya, Spin_flop, Jin} The purpose of this work is to
get more information on the magnetic phase transitions and
spin-phonon coupling of multiferroic materials by studying the
low-$T$ thermal conductivity.

Multiferroicity is a result of strong coupling between magnetic
and electric degrees of freedom in insulators and has received a
lot of research interests because of its application
usage.\cite{Tokura, Cheong} It is found that multiferroic
materials usually present complex $H-T$ phase diagrams and
multiple magnetic phase transitions, accompanied by the drastic
changes of electric properties.\cite{Tokura, Cheong, Lottermoser,
Hur, Tokunaga_DyFeO3, Tokunaga_GdFeO3} As a result, the low-$T$
heat transport may show peculiar behaviors at these transitions.
We choose GdFeO$_3$ as a candidate, which has a distorted
perovskite structure with an orthorhombic unit cell
($Pbnm$).\cite{Geller} It is known that Fe$^{3+}$ spins form an AF
order along the $a$ axis below $T_N^{Fe}$ = 661 K with a weak
ferromagnetic (WFM) component along the $c$ axis due to the spin
canting in the $ac$ plane,\cite{Treves, Bozorth} which results
from the Dzyaloshinskii-Moriya interaction.\cite{DM} The spin
structure of Fe$^{3+}$ ions can be expressed as $G_xA_yF_z$ in
Bertaut's notation,\cite{Bertaut} where $G_x$, $A_y$ and $F_z$
stand for the spin components along the $a$, $b$ and $c$ axes with
the NaCl-type, the layered-type and the ferromagnetic-type
configurations, respectively. On the other hand, Gd$^{3+}$ moments
order antiferromagnetically along the $a$ axis below $T_N^{Gd}$ =
2.5 K and show a $G_x$-type spin structure.\cite{Bertaut,
Tokunaga_GdFeO3, Cashion} The ferroelectric polarization appears
below $T_N^{Gd}$ and is considered to originate from the
spin-exchange striction;\cite{Tokunaga_GdFeO3, Tokunaga_DyFeO3,
Choi} more exactly, the interaction between adjacent Fe$^{3+}$ and
Gd$^{3+}$ layers with respective $G$-type AF
ordering\cite{Bertaut} drives Gd$^{3+}$ ions to displace along the
$c$ axis so as to induce the ferroelectric polarization along the
$c$ axis. The magnetic-field dependences of electric polarization
($P$) were carefully studied and discussed to be related to the
transitions of magnetic structures. In the case of $H \parallel
a$, the simultaneous spin-flop transition of Gd$^{3+}$ moments and
reorientation of Fe$^{3+}$ spins occur at $\sim$ 0.5
T.\cite{Tokunaga_GdFeO3, Durbin} Across this transition, the
magnetic structure changes from phase I ($G_xA_yF_z$ for Fe$^{3+}$
spins and $G_xA_y$ for Gd$^{3+}$ moments) to phase II (Fe$^{3+}$:
$F_xC_yG_z$; Gd$^{3+}$: $G_z$); thus, the electric polarization
originating from the spin exchange striction shows a sudden drop.
Upon increasing the magnetic field further, Gd$^{3+}$ spins
gradually turn to the direction of magnetic field and they are
completely polarized at $\sim$ 2 T, where the magnetic structure
changes from phase II to phase III (Fe$^{3+}$: $F_xC_yG_z$;
Gd$^{3+}$: $F_x$). In the case of $H \parallel c$, the magnetic
structure changes from phase I to phase IV (Fe$^{3+}$:
$G_xA_yF_z$; Gd$^{3+}$: $F_z$) at $\sim$ 2.5 T due to the simple
spin-polarization transition of Gd$^{3+}$
moments.\cite{Tokunaga_GdFeO3} In both cases, the electric
polarization decreases to zero as long as the Gd$^{3+}$ moments
are polarized at high field.\cite{Tokunaga_GdFeO3}

In this work, we study the low-$T$ thermal conductivity of
GdFeO$_3$ single crystals and find that the magnetothermal
conductivity is rather large, indicating a strong spin-phonon
coupling in this compound. The magnetic-field-induced spin-flop
(or spin-reorientation) transition and spin-polarization
transition are detected by $\kappa(H)$ isotherms. One peculiarity
of the heat transport of GdFeO$_3$ is that it shows a hysteresis
of the subKelvin $\kappa(H)$ curves in a broad range of magnetic
field. The possible origin of this irreversibility is discussed to
be due to the ferroelectric-domain walls scattering on phonons.
This result indicates the capability of magnetic field controlling
the ferroelectric domain structures, which is a special but
understandable phenomenon of multiferroic materials.

\section{Experiments}

High-quality GdFeO$_3$ single crystals are grown by the
floating-zone technique in a flow of oxygen.\cite{Tokunaga_GdFeO3}
The samples for thermal conductivity measurements are cut
precisely along the crystallographic axes with typical dimension
of 2.5 $\times$ 0.6 $\times$ 0.15 mm$^3$ after orientated by using
the x-ray Laue photographs. The $a$-axis and $c$-axis thermal
conductivities ($\kappa_a$ and $\kappa_c$) are measured by a
conventional steady-state technique and two different processes:
(i) using a ``one heater, two thermometers" technique in a $^3$He
refrigerator and a 14 T magnet at temperature regime of 0.3 -- 8
K; (ii) using a Chromel-Constantan thermocouple in a $^4$He
cryostat for zero-field data above 4 K.\cite{Sun_DTN, Wang} In
these measurements, the temperature gradient is 2--5 \% and $\le$
2 \% of the sample temperature for temperatures below and above 30
K, respectively. The specific heat is measured by the relaxation
method in the temperature range from 0.4 to 30 K using a
commercial physical property measurement system (PPMS, Quantum
Design).

\section{Results and Discussion}

\begin{figure}
\includegraphics[clip,width=7.0cm]{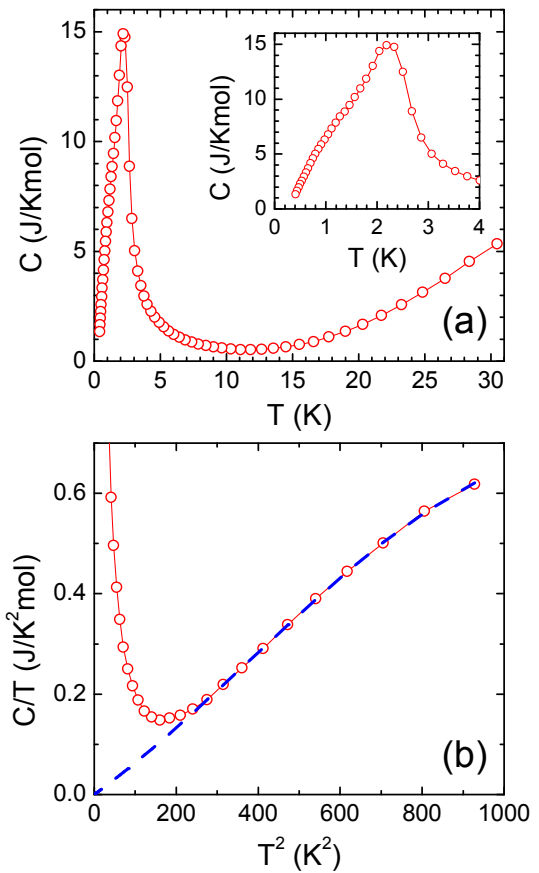}
\caption{(Color online) (a) Temperature dependence of the specific
heat of GdFeO$_3$ single crystal below 30 K. The inset displays
the data below 4 K, which show a sharp peak at 2.2 K and a
shoulder-like feature at $\sim$ 1 K. (b) The specific data plotted
in $C/T$ vs $T^2$. The dashed line shows the fitting to the
high-$T$ data by using the formula of phonon specific heat, that
is, $C = \beta T^3 + \beta_5 T^5 + \beta_7 T^7$.}
\end{figure}

Before presenting the heat transport results of GdFeO$_3$
crystals, we show in Fig. 1 the low-$T$ specific heat data. At
very low temperatures, a large peak, which is apparently of
magnetic origin, shows up. As can be seen from the inset to Fig.
1(a), this low-$T$ peak seems to consist of a sharp peak at 2.2 K
and a shoulder-like feature (or a weak peak) at $\sim$ 1 K. Note
that these data essentially reproduce those in an earlier
report.\cite{Cashion} However, there is one small difference
between two sets of data, that is, the data in Ref.
\onlinecite{Cashion} showed the sharp peak at 1.47 K and the
shoulder-like feature at $\sim$ 2 K. It was discussed that those
two features are originated from the N\'{e}el transition of
Gd$^{3+}$ moments and a Schottky contribution, respectively. So
the data in Fig. 1 indicate that the N\'{e}el temperature of
Gd$^{3+}$ moments of our GdFeO$_3$ crystal is 2.2 K, close to the
value of a recent report by a susceptibility
measurement.\cite{Tokunaga_GdFeO3}

It can be seen that the magnetic contributions to the specific
heat are important only at very low temperatures and are likely to
be negligible above $\sim$ 12 K, where the data in Fig. 1(a) shows
a minimum. So one can make an estimation of phonon specific heat
from the high-$T$ data in Fig. 1. It is known that in the
temperature range 0.02 $< T / \theta_D <$ 0.1 ($\theta_D$ is the
Debye temperature), one had better use the low-frequency expansion
of the Debye function, $C = \beta T^3 + \beta_5 T^5+\beta_7 T^7 +
...$, where $\beta$, $\beta_5$ and $\beta_7$ are
temperature-independent coefficients.\cite{Tari} It is found that
this formula gives a precise fitting to the experimental data
above 15 K, as shown in Fig. 1(b), with the fitting parameters
$\beta = 5.90 \times 10^{-4}$ J/K$^4$mol, $\beta_5 = 4.52 \times
10^{-7}$ J/K$^6$mol and $\beta_7 = -3.96 \times 10^{-10}$
J/K$^8$mol. Note that at very low temperatures, the $T^5$- and
$T^7$- terms are negligible and the phonon specific heat shows a
well-known $T^3$ dependence with the coefficient of $\beta$.

\begin{figure}
\includegraphics[clip,width=7.5cm]{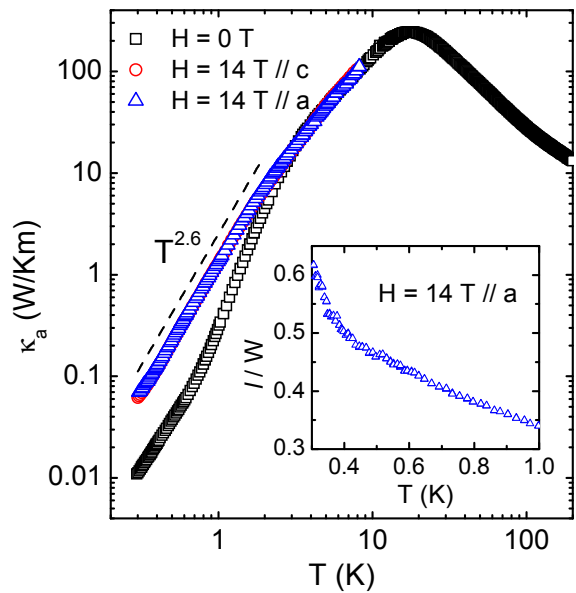}
\caption{(Color online) Temperature dependences of the $a$-axis
thermal conductivity of GdFeO$_3$ single crystal in zero and 14 T
magnetic field parallel to the $a$ and $c$ axis. The data in 14 T
field for two directions are nearly coincident. The dashed line
indicates the $T^{2.6}$ dependence. The inset shows the
temperature dependence of the phonon mean free path $l$ divided by
the averaged sample width $W$ in 14 T magnetic field.}
\end{figure}

Figure 2 shows the temperature dependences of $\kappa_a$ for
GdFeO$_3$ single crystals in zero and 14 T magnetic field parallel
to the $a$ and $c$ axis. As far as the zero-field data are
concerned, they show a typical phonon transport behavior at
relatively high temperatures. \cite{Berman} In particular, the
magnitude of the phonon peak at 18 K is as large as 250 W/Km,
which is rather rare in the transition-metal oxides and indicates
weak crystal defects or impurities. At low temperatures, however,
there are some features showing a complexity of the phonon
transport in this materials. First, even at very low temperatures
the $\kappa_a(T)$ data show a distinct deviation from the $T^3$
law, a sign of the phonon boundary-scattering limit, which
apparently indicates the remaining of significant microscopic
scattering of phonons.\cite{Berman} Actually, the zero-field curve
exhibits a weak kink-like temperature dependence below 2 K, which
is likely due to the magnon-phonon scattering since the Gd$^{3+}$
moments order antiferromagnetically below 2.2
K.\cite{Tokunaga_GdFeO3, Vitebskii} The effect of magnetic field
on thermal conductivity seems to confirm this possibility. When 14
T magnetic field is applied, the conductivities below 2 K become
larger and the kink disappears, which clearly indicates the
negative effect of magnons on the heat transport, considering that
at low temperatures the magnons can hardly be thermally excited in
high field. A $T^{2.6}$ dependence, which is very close to the
boundary scattering limit, indicates that magnetic scattering on
phonons is almost smeared out in 14 T field. In addition, the
effect of strong magnetic field on $\kappa_a$ is essentially
isotropic for $H \parallel a$ and $H \parallel c$.

It is possible to estimate the mean free path of phonons at low
temperatures and to judge whether the phonons are free from
microscopic scattering at subKelvin temperatures. The phononic
thermal conductivity can be expressed by the kinetic formula
$\kappa_{ph} = \frac{1}{3}Cv_pl$,\cite{Berman} where $C = \beta
T^3$ is phonon specific heat at low temperatures, $v_p$ is the
average velocity and $l$ is the mean free path of phonon. Here
$\beta = 5.90 \times 10^{-4}$ J/K$^4$mol is obtained from the
above specific-heat data and $v_p$ = 1930 m/s can be estimated
from Deybe temperature $\Theta_D$ using the relations $\beta =
\frac{12\pi^4}{5} \frac{Rs}{\Theta_D^3}$ and $\Theta_D =
\frac{\hbar v_p}{k_B} (\frac{6\pi^2
Ns}{V})^\frac{1}{3}$,\cite{Tari} where $N$ is the number of
molecules per mole and each molecule comprises $s$ atoms, $V$ is
the volume of crystal and $R$ the universal gas constant. So we
can calculate $l$ from the 14 T $\kappa(T)$ data and compare it
with the averaged sample width $W = 2\sqrt{A/\pi}$ = 0.361
mm,\cite{Berman, Sun_Comment} where $A$ is the area of cross
section. As shown in the inset to Fig. 2, the ratio $l / W$
increases with lowering temperature and becomes close to one at
0.3 K, which means that the boundary scattering limit is nearly
established at such low temperatures. On the other hand, although
not shown in the figure, the mean free path of phonons in zero
field is apparently several times smaller than that in 14 T field,
which obviously demonstrates the significance of microscopic
phonon scattering in zero field.

\begin{figure}
\includegraphics[clip,width=8.5cm]{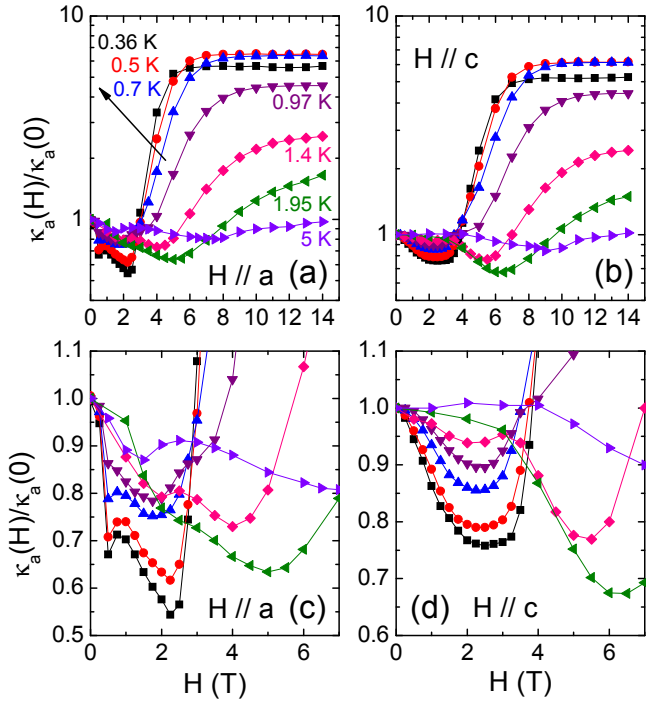}
\caption{(Color online) (a,b) Magnetic-field dependences of the
$a$-axis thermal conductivity of GdFeO$_3$ single crystal. The
magnetic fields are applied along the $a$ or $c$ axis. All the
data are measured in the field sweeping-up process after
zero-field cooling. (c,d) Zoom in of the low-field data of panels
(a) and (b).}
\end{figure}

The magnon-phonon scattering is further evidenced by the
magnetic-field dependences of $\kappa_a$ at low temperatures, as
shown in Figs. 3(a) and 3(b). For both $H \parallel a$ and $H
\parallel c$, the $\kappa_a(H)$ isotherms show a reduction at low
fields followed by an enhancement at high fields. At subKelvin
temperatures, $\kappa_a$ achieves a saturated value in high
magnetic fields, but the magnitude weakens gradually with
increasing temperature. Note that the high-field enhancement of
thermal conductivity can be as high as $\sim$ 700\%, while the
strongest suppression at low fields is $\sim$ 50\%, which clearly
indicates that the coupling between magnons and phonons is pretty
strong.\cite{Wang} Another remarkable feature is that at subKelvin
temperatures there are two ``dips" at low fields ($<$ 4 T) for $H
\parallel a$, while only one shallow and broad ``dip" emerges for
$H \parallel c$. Moreover, the field-induced suppression of
conductivity becomes weaker as increasing temperature. At higher
temperatures above 1 K, another ``dip" appears at high fields ($>$
4 T) for both cases and it becomes broader and shifts to higher
field with increasing temperature, which can be clearly seen in
Figs. 3(c) and 3(d), suggesting its possible origin from phonon
scattering by paramagnetic moments.\cite{Berman, Sun_PLCO,
Sun_GBCO} We have also measured the $c$-axis thermal conductivity
for both magnetic-field directions (see the Appendix), and found
that the behavior of $\kappa$ is essentially dependent on the
direction of magnetic field rather than the direction of heat
current, which again demonstrates the role of magnons in the heat
transport in magnetic field.

Figures 3(c) and 3(d) show the details of low-field $\kappa_a(H)$
behaviors. As mentioned above, there are clear differences in the
$\kappa_a(H)$ isotherms between $H \parallel a$ and $H \parallel
c$. When the magnetic field is parallel to the $a$ axis, the two
``dips" of $\kappa_a(H)$ locate at $\sim$ 0.5 T and $\sim$ 2.25 T,
which are weakly temperature dependent. Apparently, these two
``dips" are directly related to the spin-flop transition and the
spin polarization of Gd$^{3+}$ magnetic structure,\cite{Spin_flop}
as suggested by the low-$T$ electric polarization
data.\cite{Tokunaga_GdFeO3} It was discussed that at 0.5 T the
magnetic structure changes from phase I to phase II, which is
associated with a simultaneous occurrence of the Fe$^{3+}$ spin
reorientation and the Gd$^{3+}$ spin flop both from mainly along
the $a$ axis to mostly along the $c$ axis. In Bertaut's notation,
the magnetic structure changes from $G_xA_yF_z$ to $F_xC_yG_z$ for
the Fe$^{3+}$ spins and from $G_xA_y$ to $G_z$ for the Gd$^{3+}$
moments, respectively. With increasing magnetic field further, the
Gd$^{3+}$ moments gradually rotate to the direction of magnetic
field and are fully polarized at $\sim$ 2.25 T, where the magnetic
structure changes to phase III (represented as $F_xC_yG_z$ for
Fe$^{3+}$ spins and $F_x$ for Gd$^{3+}$ moments). Since the magnon
excitations become gapless at either the spin-flop field or the
spin-polarization field, the quickly increased number of magnons
can scatter phonons strongly and cause a drastic decrease of
thermal conductivity,\cite{Spin_flop, Jin, Wang} manifested as
those two weakly-temperature-dependent ``dips" in $\kappa(H)$
curves. When the magnetic field is applied along the $c$ axis,
there is no spin-flop transition of the magnetic structure;
instead, the Gd$^{3+}$ ions change their spin direction under a
simple spin polarization process with increasing
field.\cite{Tokunaga_GdFeO3} It was discussed that the Gd$^{3+}$
ion turns spin direction gradually from almost $a$ axis to
complete $c$ axis at $\sim$ 2.5 T, and the magnetic structure
changes from phase I to phase IV ($G_xA_yF_z$ for Fe$^{3+}$ spins
and $F_z$ for Gd$^{3+}$ moments).\cite{Tokunaga_GdFeO3}
Consequently, there is only one broad ``dip" in $\kappa_a(H)$ for
$H \parallel c$.

\begin{figure}
\includegraphics[clip,width=8.5cm]{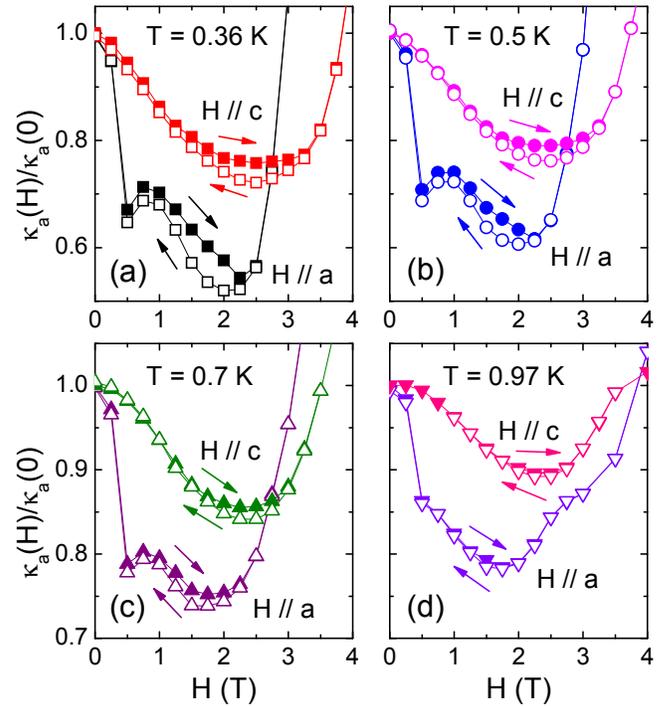}
\caption{(Color online) Low-field $\kappa_a(H)$ isotherms of
GdFeO$_3$ single crystal at subKelvin temperatures with magnetic
field along the $a$ and $c$ axis. The data shown with solid
symbols are measured in the ascending field after the sample is
cooled in zero field, while the open symbols show the data with
the descending field, as indicated by arrows.}
\end{figure}

Therefore, all the above results can be well understood in the
scenario of phonon heat transport, with the significant scattering
by magnetic excitations. However, this is not the whole story. It
is found that the low-$T$ thermal conductivity of GdFeO$_3$ is
dependent on the history of applying magnetic field.

As shown in Fig. 4, for both $H \parallel a$ and $H \parallel c$,
the $\kappa_a(H)$ isotherms measured with field increasing are
larger than those with field decreasing, forming a clear
hysteresis at low temperatures. This phenomenon demonstrates that
at very low temperatures, there is some peculiar channel of phonon
scattering that is related not only to the magnetic field but also
to the history of applying field. As discussed on Figs. 2 and 3,
the magnons are effective phonon scatterers in zero and low
fields. However, they cannot simply produce an irreversible
behavior of $\kappa(H)$ in such a broad field range from almost
zero to $\sim$ 3 T. In particular, the spin polarization is a
naturally continuous transition and the hysteretic behavior cannot
be expected across this transition. On the other hand, other
microscopic phonon scatterers like point defects or dislocations
in crystal structure are also irrelevant for two
reasons.\cite{Berman} First, the scattering processes from crystal
imperfections have nothing to do with the external magnetic field,
let alone the history of applying field; second, these scatterings
are known to be less effective upon lowering temperature, but the
hysteresis becomes more pronounced with decreasing temperature.

Then, what else can be the origin of this kind of scattering that
produces the hysteresis? Considering the development of hysteresis
upon lowering temperature, a natural origin is related to the
magnetic or ferroelectric domains in this multiferroic material,
in which the domain walls can play an important role in scattering
phonons at very low temperatures when the mean free path of
phonons is long enough to be comparable to the interval of domain
walls. It is known from the former work that the
weak-ferromagnetism-related domains in GdFeO$_3$ can be produced
only at very low field ($<$ 0.1 T).\cite{Tokunaga_GdFeO3}
Therefore, the ferroelectric domain walls are likely the main
source of phonon scattering that is responsible for the
$\kappa(H)$ irreversibility in the broad field range well above
0.1 T. In this regard, there are several experimental results
supporting the origin of $\kappa(H)$ hysteresis from the
ferroelectric domains. First, the electric polarization vs
magnetic field $P(H)$ also showed the irreversible behaviors in
the absence of electric field for both $H \parallel a$ and $H
\parallel c$.\cite{Tokunaga_GdFeO3} Second, the field range that
the hysteretic behavior of $\kappa(H)$ appears is almost the same
as that exhibiting the hysteretic $P(H)$
curves.\cite{Tokunaga_GdFeO3} Third, the relative magnitude of
polarization for field increasing and field decreasing indicated
that there are less ferroelectric domains when the field is
increasing.\cite{Tokunaga_GdFeO3} It is therefore in good
agreement with the present observation that the thermal
conductivity is larger in the field-increasing branch, which can
be due to the weaker phonon scattering by domain walls.
Furthermore, with increasing temperature the size of hysteresis
diminishes and disappears completely above $\sim$ 1 K, also in
good agreement with the fact that the phonon scattering by
boundaries are unimportant at high temperatures where the mean
free path of phonons becomes much shorter than the averaged
distance of domain walls.

\begin{figure}
\includegraphics[clip,width=8.5cm]{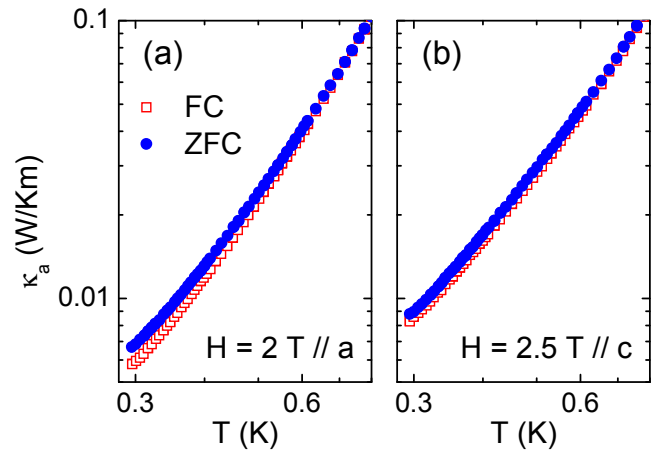}
\caption{(Color online) Temperature dependences of the $a$-axis
thermal conductivity of GdFeO$_3$ single crystal in 2 T magnetic
field along the $a$ axis (a) and in 2.5 T field along the $c$ axis
(b). The data are taken with slowly warming up the sample from the
lowest temperature after field cooling or zero-field cooling.}
\end{figure}

The magnetic-field-induced irreversibility of $\kappa$ can also be
manifested in the $\kappa(T)$ curves, as shown in Fig. 5, measured
in some characteristic fields by warming up the sample after
zero-field cooling or field cooling to the lowest temperature. It
is found that below 1 K the zero-field cooled (ZFC) conductivities
are larger than the field cooled (FC) ones, which coincides with
the hysteresis of $\kappa(H)$ curves. This difference indicates
that there are less ferroelectric domains if the sample is cooled
in zero field. From these data, one can get a rough estimation of
the averaged distance between the ferroelectric domain walls,
which usually has the same order of magnitude to that of the mean
free path of phonons as long as the domain walls are effective
phonon scatterers. Following the above calculation on the 14 T
$\kappa(T)$ data, the mean free path of phonons is obtained to be
20--30 $\mu$m from the ZFC data in Fig. 5. This is a rather
reasonable value of domain size for the ferroelectric single
crystals.

Note that the hysteresis of $\kappa(H)$ of a magnetic material
itself is not strange. The irreversibility has been known to often
appear at the field-induced first-order magnetic transition, such
as the spin-Peierls to AF order\cite{Takeya} and the
liquid-gas-like transition in spin-ice compounds,\cite{Sun_DTO}
etc. However, the hysteretic $\kappa(H)$ of GdFeO$_3$ is somewhat
different, because it appears in a rather broad field range where
the magnetization does not show
irreversibility.\cite{Tokunaga_GdFeO3} In contrast, another
well-known large hysteresis of low-$T$ $\kappa(H)$ is the one
observed in the high-$T_c$ cuprate Bi$_2$Sr$_2$CaCu$_2$O$_8$
(BSCCO),\cite{Aubin1, Aubin2} in which the hysteresis is well
understood by the vortex-pinning effect that is accompanied with
the irreversible macroscopic magnetization. From the above
discussions, one can see that the irreversible $\kappa(H)$ of
GdFeO$_3$ is most likely caused by the ferroelectric domain walls
scattering on phonons, that is dependent on the history of
applying the magnetic field. It should be pointed out that the
phonon scattering by the ferroelectric domain walls had been
studied by modifying the domain structure through applying the
electric field directly.\cite{Weilert, Steigmeier} It was proved
in KH$_2$PO$_4$ that the low-$T$ thermal conductivity of the
single-domain state is much larger than that of the multi-domain
state.\cite{Weilert} Similar result was obtained in SrTiO$_3$, in
which the low-$T$ thermal conductivity under high electric field
is clearly larger than that in zero field.\cite{Steigmeier}
However, these former experiments should be analyzed very
carefully because applying an electric field using the contacts on
the sample surface could bring some uncertainties in the thermal
conductivity measurements. In this regard, changing the domain
structures by applying magnetic field is free from such problem
and can give more reliable data. It is intriguing that in
multiferroic GdFeO$_3$ the ferroelectric domain structures are
able to be manipulated by the magnetic field, which is a key to
make the peculiar low-$T$ $\kappa(H)$ hysteresis to be observable.

\section{Summary}

The low-temperature heat transport of GdFeO$_3$ single crystal is
found to be strongly dependent on the magnetic field and is
helpful for studying the magnetic transitions and magnetoelectric
coupling in this material. The magnons play an important role in
the low-$T$ thermal conductivity by scattering phonons instead of
acting as heat carriers. As a result, both the spin-flop or
reorientation of the magnetic structure and spin polarization
cause significant phonon scattering at the transition fields. The
most remarkable result is that the low-$T$ ($<$ 1 K) thermal
conductivity shows an irreversible behavior on the history of
applying magnetic field. This phenomenon is rather peculiar in the
sense that it appears in a broad range of magnetic field ($\sim$
0--3 T) where the magnetization is known to be almost reversible.
Furthermore, the hysteresis of $\kappa(H)$ or the irreversibility
of ZFC and FC $\kappa(T)$ become larger upon lowering
temperatures. All these data suggest that the ferroelectric domain
structures, which are surprisingly sensitive to the history of
applying magnetic field, are playing the key role in producing the
irreversibility of thermal conductivity through the phonon
scattering by the domain walls. The present results show an
interesting case that the ferroelectric domains of a multiferroic
material can be manipulated by magnetic field, and this kind of
magnetoelectric coupling can be observed by the heat transport
measurement at very low temperatures.

\begin{acknowledgements}

We thank A. N. Lavrov and Y. Tokunaga for helpful discussions.
This work was supported by the Chinese Academy of Sciences, the
National Natural Science Foundation of China, the National Basic
Research Program of China (Grants No. 2009CB929502 and No.
2011CBA00111), and the RFDP (Grant No. 20070358076).

\end{acknowledgements}

\section*{Appendix: The $c$-axis thermal conductivity of GdFeO$_3$}

\begin{figure}
\includegraphics[clip,width=7.5cm]{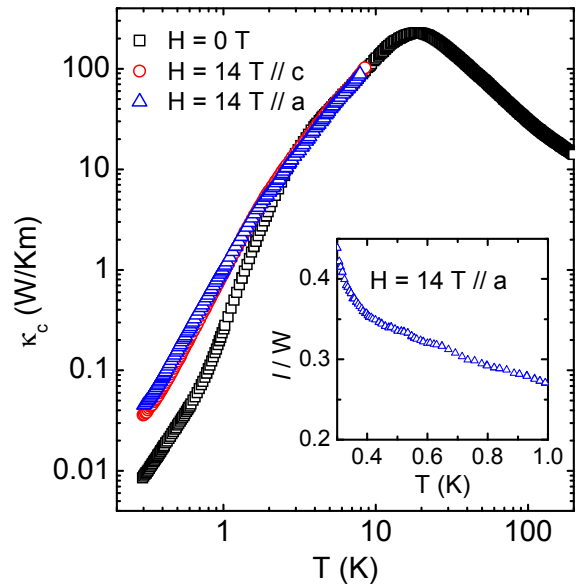}
\caption{(Color online) Temperature dependences of the $c$-axis
thermal conductivity of GdFeO$_3$ single crystal in zero and 14 T
magnetic field along the $a$ or $c$ axis. The inset shows the
temperature dependence of the phonon mean free path $l$ divided by
the averaged sample width $W$ (= 0.344 mm) in 14 T magnetic
field.}
\end{figure}

Figure 6 shows the temperature dependences of $\kappa_c$ for
GdFeO$_3$ single crystals in zero and 14 T magnetic field parallel
to the $a$ and $c$ axis. Apparently, these data are essentially
the same as those of $\kappa_a(T)$, including all the main
features like the magnitude of phonon peak, the weak-kink-like
temperature dependence below $\sim$ 2 K, the significant recovery
of conductivity in 14 T field, etc. In addition, the temperature
dependences of 14 T thermal conductivities are also similar to
those of $\kappa_a(T)$ in 14 T. The calculated mean free path of
phonons along the $c$ axis is found to have the same order of
magnitude to that from $\kappa_a$, as shown in the inset to Fig.
6. One can easily conclude that the phonon heat transport is
nearly isotropic in GdFeO$_3$.

\begin{figure}
\includegraphics[clip,width=8.5cm]{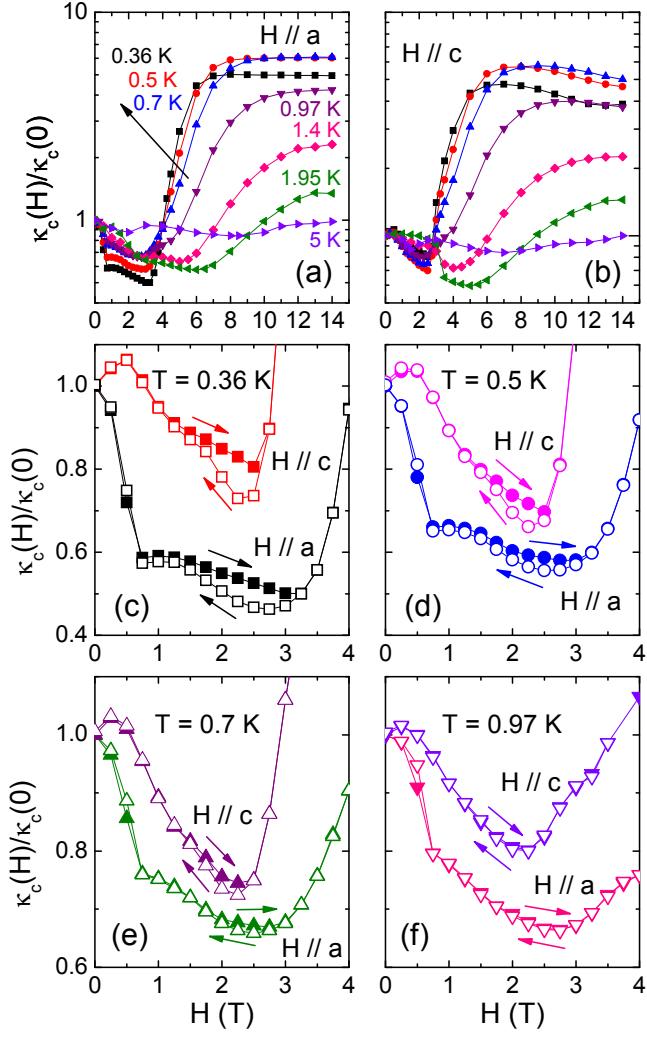}
\caption{(Color online) (a,b) Magnetic-field dependences of the
$c$-axis thermal conductivity of GdFeO$_3$ single crystal. The
magnetic fields are applied along the $a$ or $c$ axis and all the
data are measured in the field sweeping-up process after
zero-field cooling. (c--f) Low-field $\kappa_c(H)$ isotherms of
GdFeO$_3$ single crystal at subKelvin temperatures with magnetic
field along the $a$ and $c$ axis. The solid symbols show data
measured with ascending field after the sample is cooled in zero
field, while the open symbols show the data with descending field,
as indicated by arrows.}
\end{figure}

Figures 7(a) and 7(b) show the magnetic-field dependence of the
$c$-axis thermal conductivity of GdFeO$_3$ single crystal. The
overall behaviors of $\kappa_c(H)$ are qualitatively the same as
those of $\kappa_a(H)$, that is, the thermal conductivity are
suppressed at low fields and strongly enhanced at high fields.
Moreover, at low fields there are also two ``dips" for $H
\parallel a$ and one ``dip" for $H \parallel c$, respectively. One
may note that the ``dip" fields of $\kappa_c$ for both $H
\parallel a$ and $H \parallel c$ are somewhat different from those
of $\kappa_a$ when the field is applied along the same direction.
However, such discrepancy is likely due to the demagnetization
effect. Both samples have size about 2.5 $\times$ 0.6 $\times$
0.15 mm$^3$. For such kind of long-shaped samples, the
demagnetization factor $n$ is negligible ($\approx 0$) when the
applied field is along the longest dimension, while it is not
negligible (taking a value between 0 and 1) when the field is
along the shortest dimension. For the $a$-axis sample, its $a$
axis is along the longest dimension and the $c$ axis along the
shortest dimension; while the $c$-axis sample has the $c$ axis and
$a$ axis along the longest and the shortest dimensions,
respectively. Therefore, for the experimental configurations of
($\kappa_a$, $H \parallel a$) and ($\kappa_c$, $H \parallel c$),
the inner magnetic field ($H_i$) is equal to the external field
($H$) since $n \approx 0$; while for the configurations of
($\kappa_a$, $H \parallel c$) and ($\kappa_c$, $H \parallel a$),
the inner field is different to the external one and can be
expressed as $H_i = H/(1+n\chi)$, where $\chi$ is the magnetic
susceptibility. Note that although the factor $(1+n\chi)$ is not
available for our samples, it is possible to make a comparison
between $\kappa_a(H)$ and $\kappa_c(H)$ data with taking into
account the demagnetization effect. The comparison of the ``dip"
fields between $\kappa_a(H)$ and $\kappa_c(H)$ for both $H
\parallel a$ and $H \parallel c$ indicated that the factor
$(1+n\chi)$ is about 1.5. For example, in the case of ($\kappa_a$,
$H \parallel a$) the ``dip" fields of 0.5 T and 2.25 T are
intrinsic values of transition fields; while the ``dip" fields of
0.75 T and 3.25 T in the case of ($\kappa_c$, $H \parallel a$)
re-scaled by the factor of 1.5 give values of 0.5 T and 2.2 T,
respectively, which both match the ``dip" fields of $\kappa_a(H)$
very well. Under such kind of consideration, it is a bit strange
that the ``dip" fields of ($\kappa_a$, $H \parallel c$) and
($\kappa_c$, $H \parallel c$) do not differ much from each other.
Therefore, the quantitative analysis of these data are called for.

The $\kappa_c(H)$ isotherms for both $H \parallel a$ and $H
\parallel c$ also present the hysteresis on the magnetic field
sweeping up and down, as shown in Figs. 7(c--f). It can be seen
that the field region and temperature regime for showing the
hysteresis, the difference between the field up and down data are
all very similar to those of $\kappa_a(H)$ data. Apparently, these
results have the same origin as those of $\kappa_a(H)$.

\begin{figure}
\includegraphics[clip,width=8.5cm]{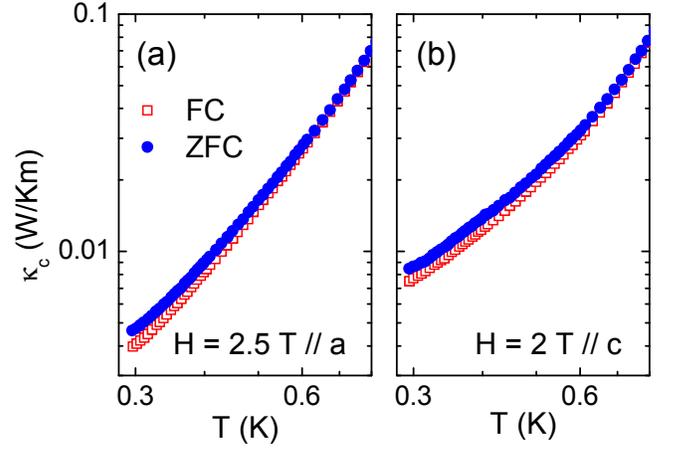}
\caption{(Color online) Temperature dependences of the $c$-axis
thermal conductivity of GdFeO$_3$ single crystal in 2.5 T magnetic
field along the $a$ axis (a) and in 2 T field along the $c$ axis
(b). The data are taken with slowly warming up the sample from the
lowest temperature after field cooling or zero-field cooling.}
\end{figure}

Figure 8 shows the magnetic-field-induced irreversibility of
$\kappa_c(T)$, measured in some characteristic fields by warming
up the sample after zero-field cooling or field cooling to the
lowest temperature. They also show similar behaviors to those of
the ZFC and FC $\kappa_a(T)$ data.

In summary, the low-$T$ heat transport of GdFeO$_3$ is nearly
isotropic. There are strong coupling between the phonons and
magnons, which determines the main features of the magnetic-field
dependence of thermal conductivity. The characteristic transitions
of $\kappa(H)$ isotherms are strongly dependent on the direction
of external field and are caused by the field-induced transitions
of magnetic structure. An irreversibility of low-$T$ thermal
conductivity on the magnetic field is observed and is likely due
to the phonon scattering by the ferroelectric domain walls. The
peculiarity is that this phenomenon points to the manipulation of
ferroelectric properties by magnetic field, which is a
characteristic of multiferroic materials.

\end{document}